\documentstyle[10pt]{article}
\begin{document}
\title{Stationary states of an electron in periodic
structures  in a constant uniform electrical field}
\author{N. L. Chuprikov \\ Tomsk State University, Tomsk, RUSSIA\\
E-mail: chnl@phys.tsu.tomsk.su}
\maketitle
\begin{abstract}
On the basis of the transfer matrix technique an analytical method to
investigate the stationary states, for an electron in one-dimensional
periodic structures in an external electrical field, displaying the
symmetry of the problem is developed.  These solutions are shown to be
current-carrying. It is also shown that the electron spectrum for
infinite structures is continuous, and the corresponding wave functions
do not satisfy the symmetry condition of the problem.
\par
\noindent Keywords: Wannier-Stark problem, Zener tunneling,
Bloch oscillates
\end{abstract}
\newpage
\section{Introduction}

The question of how an external (constant, uniform) electrical field
influences the electron motion in periodic structures has been of
great interest for decades \cite{Blo,Zen,Wan,Zak}.
Nevertheless, the disagreements on the nature of the energy spectrum
still persist at present. Some analytical investigations
\cite{Wan,Kri,Ro2} show that the energy spectrum should be discrete
irrespective of the potential form and consist of the so-called
Wannier-Stark ladders with uniformly spaced levels. But other works
(see \cite{Zak,Ra2,Ao1,Av2} and references therein, including
rigorous mathematical results for the smooth potentials) point to the
fact that under certain restrictions on the potential the spectrum is
continuous, and a discrete spectrum may exist only for the periodic
structures consisting of the $\delta$ --- potentials (under certain
conditions) and $\delta^\prime$ --- potentials (always).

In the simplest model the problem is reduced to solving
the one-dimensional Schr\"{o}dinger equation whose Hamiltonian
includes the periodic potential and the potential of an electric field.
It is known \cite{Kri} that the properties of the equation depend
strongly on the choice of a gauge for the field. When a scalar
potential is used, the Hamiltonian is time independent, as in the
absence of the field (the problem with the zero field will be referred
to as a zero-field problem (ZFP)).  But its symmetry is different from
the translational one.  In this case it is important to reveal
the changes, in the band-gap energy spectrum of the ZFP,  caused by the
field and to find the wave functions satisfying the new conditions of
symmetry (these functions will play the role which is similar to that
of the Bloch functions in the ZFP). A directly opposite situation
arises for a vector representation. Now, switching on the field does
not break the translational symmetry, and the Hamiltonian becomes time
dependent.  As a result, the electron energy is no longer a quantum
number and the initial problem can be treated as the one on the Bloch
electron accelerated by the field.

The mathematical difficulties associated with making use of the scalar
potential are well known beginning with the famous paper \cite{Wan} by
Wannier.  To overcome them, the author had to treat a finite number
of the Bloch bands.  This approximation was rightly disputed later
\cite{Zak,Ra2}.  Recent investigations (see for example \cite{Ao1,Av2})
show that the solution of the problem essentially depends on the
alternation order of Bloch bands and gaps in the high energy region in
the spectrum of the ZFP.  It is known that for periodic finite-value
potentials the band width increases to infinity with increasing energy,
while the gap width vanishes. Taking into account a finite number of
Bloch bands is equivalent to the fact that the whole high-energy region
is a gap. Such an approximation is sure to result in a discrete
spectrum.

As far as we know, the rigorous analytical solution to the problem with
the scalar potential of the general form have not been found.  Besides,
the stationary electron states, displaying the symmetry of the problem,
remain to be investigated. In this work we propose an exact analytical
method to find such stationary states. The connection between them and
the Bloch states is discussed here. The energy spectrum of an electron
and the Zener tunneling are also considered.

\section {Symmetry of wave functions}

The basis for our approach is the transfer matrix method (TMM)
\cite{Ch1}, that we have used \cite{Ch2} for solving the ZFP.
We shall remind that one of the main points of that formalism is the
notion of out-of-barrier regions (OBR), where the total potential is
equal to zero. Here we shall use this notion as well, having made
the necessary generalizations with reference to the problem at hand. It
can be made in two ways. Firstly, one may consider that the
total potential in the OBR coincides with the Stark potential which is
a linear function of $x$ (in this case the treatment should be based on
the Airy-functions formalism).  Secondly, one may consider that the
potential in these regions is a constant which depends linearly on the
cell number.  The proportionality coefficient depends on the
electrical-field strength.  Both the variants can be used in our
approach.  However in this work we dwell on the latter because there is
a more evident association with the ZFP in this case.

The stationary Schr\"{o}dinger equation for a structure of $N$
periods (unit cells) may be written as

\begin{equation} \label{1}
\frac{d^2\Psi}{dx^2}+\frac{2m}{\hbar^2}
\bigl(E-V(x)\bigr)\Psi=0,
\end{equation}

\noindent where $E$ is the electron energy; $m$ is its mass;
$V(x)$ is defined by expressions: $V(x)=v(x)-n\Delta$, if $x\in
(a_n,b_{n+1})$ ($n=0,\ldots,N-1$); $V(x)=-n\Delta$, if $x\in (b_n,
a_n)$; $b_n=nD$; $a_n=l+nD$ ($n=0,\ldots,N$); $\Delta=e{\cal E}D$; $e$
is the electron charge (by modulus); $l$ is the OBR width; $D$ is the
structure period; $\cal E$ is the electric-field strength; $v(x)$ is a
bounded $D$-periodical function.

It should be noted that boundary conditions at the points $x=0$ and
$x=a_N$ are not needed here, for we do not solve the boundary-value
problem. Semi-infinite and infinite structures will be considered
below. Notice also that the parameter $l$ may be equal to zero. For the
OBR always may be included into the initial potential (as the point
with an infinitesimal vicinity) without changing  the solution of Eq.
(\ref{1}) (the new and initial potentials are equivalent functions).
Thus the present method can be used for any initial potential.

\newcommand{\dff}{$\Psi_{\cal E}(x;E)$ }

It is known \cite{Wan} that if the function $\Psi_E(x)$ is a solution
of Eq. (\ref{1}), then $\Psi_{E+\Delta}(x-D)$ is a solution too. On the
basis of this statement one can assume that there are solutions (to be
referred to as \dff) among those of Eq. (\ref{1}), satisfying the
condition

\begin{equation} \label{2}
\Psi_{\cal E}(x+D;E)=const\cdot\Psi_{\cal E}(x;E+\Delta),
\end{equation}
where $const$ is a complex value. Our main goal is to find the
functions \dff and to examine their properties.

The solutions to the Schr\"{o}dinger equation, both with the scalar and
with the vector potentials, are generally found in the form of an
expansion in orthogonal functions (for example, in the Bloch
functions).  In this case the required and basis functions are supposed
to belong to the same class of functions. The disadvantage of such an
approach is that, in finding wave functions obeying the symmetry
conditions, it is not always clear to what class of functions they are
to belong. So, for example, in the absence of the field the same
condition of the translational symmetry (it coincides with (\ref{2})
when $\Delta=0$) leads, in the bands, to the Bloch functions bounded
everywhere, but in the gaps it provides the functions unbounded at the
plus or minus infinity. As will be shown further, the functions \dff are
unbounded when $x\to -\infty$.  Thus, if we attempted to find these
functions in the Bloch or Wannier expansions, we could obtain an
incorrect result, because both sets of functions belong to other
classes. The transfer matrix method is free from this demerit, for
the expansions technique is not used there.

\section {Functional equation for wave functions  in the transfer
matrix method}

\newcommand{\Av}{{\cal A}}

The general solution of the Eq. (\ref{1}) in the OBR's is

\begin{equation} \label{3}
\Psi(x;E)=A_n^{(+)}(E)\exp[ik_n(x-b_n)]+A_n^{(-)}(E)
\exp[-ik_n(x-b_n)], \end{equation}

\noindent where $k_n=\sqrt{2m(E+n\Delta)/\hbar^2}$; $n=0,\ldots,N$.

Here the main problem is to find the coefficients $A_n^{(+)}(E)$ and
$A_n^{(-)}(E)$; $n=0,\ldots,N$. Once the coefficients have been
found, the determination of the \dff in the barrier regions should
present no principal problems. In the general case for this purpose one
can use, for example, the numerical technique \cite{Ch3}.

The connection between the coefficients of the solution in the first
two OBR's is given by

\begin{equation} \label{4}
\Av_0(E)=\alpha(E) Y(E)\Gamma(E)\Av_1(E);\end{equation}

Here $Y$ is a transfer matrix (see \cite{Ch1}), describing the barrier
at the unit cell $n=0$ (providing that there is no step at the
point $b_1$), and $\alpha\cdot\Gamma$ is a matrix matching the
solutions at the step at $x=b_1$:

\begin{equation} \label{5}
Y=\left(\begin{array}{cc}\tilde{q} & \tilde{p} \\\tilde{p}^* &
\tilde{q}^* \end{array} \right), \hspace{8mm}
\Gamma=\left(\begin{array}{cc}q_s & p_s \\p_s & q_s
\end{array} \right); \hspace{8mm}
\Av_n=\left(\begin{array}{c} A_n^{(+)} \\ A_n^{(-)}
\end{array} \right);
\end{equation}

\[
\tilde{q}=\frac{1}{\sqrt{T}}\exp[-i(J+k_0l)]; \hspace{8mm}
\tilde{p}=\sqrt{\frac{R}{T}}\exp[i(\frac{\pi}{2}+F-k_0l)];
\]
\[
q_s=(\alpha+\alpha^{-1})/2;
\hspace{8mm} p_s=(\alpha^{-1}-\alpha)/2; \hspace{8mm}
\alpha(E)=\sqrt{k_1(E)/k_0(E)};
\]
the phases $J(E)$, $F(E)$ and the transmission coefficient $T(E)$ (see
\cite{Ch1}) describing the barrier in the zero cell are supposed to
be known; $R=1-T$.

Let
\[Z=Y\Gamma =\left(\begin{array}{cc}q & p
\\p^* & q^* \end{array} \right), \]
then the relationship (\ref{4}) can be rewritten as

\begin{equation} \label{8}
\Av_0(E)=\alpha(E)Z(E)\Av_1(E),
\end{equation}
and the connection between any two adjacent OBR's will be
determined by

\begin{equation} \label{9}
\Av_n(E)=\alpha(E+n\Delta)Z(E+n\Delta)\Av_{n+1}(E); \hspace{8mm}
n=0,1,\ldots,N-1.
\end{equation}
Providing (\ref{9}) the connection between the zero and $N$-th unit cells
can be written in the form

\begin{equation} \label{10}
\Av_0(E)=\alpha_{(1,N)}(E) {\cal Z}_{(1,N)}(E)\Av_N(E),
\end{equation}
where
\begin{equation} \label{11}
{\cal Z}_{(1,N)}(E)=Z(E)\cdot\ldots
\cdot Z(E+(N-1)\Delta);
\end{equation}

\[\alpha_{(1,N)}(E)=\prod_{n=0}^{N-1}\alpha(E+n\Delta)=
\sqrt{\frac{k_0(E+N\Delta)}{k_0(E)}}.\]

Defining for all $n$ the vector

\[\tilde{\Av}_n(E)=\alpha_{(1,n)}(E)\Av_n(E),\]
we can rewrite Eq. (\ref{10}) as

\begin{equation} \label{12}
\Av_0(E)\equiv\tilde{\Av}_0(E)={\cal Z}_{(1,N)}(E)\tilde{\Av}_N(E).
\end{equation}

Now, by analogy with the ZFP \cite{Ch2} we will attempt to find the
wave functions whose expressions for the extreme OBR's (i.e.  for the
zero and $N$-th unit cells) are connected by means of symmetry. For this
purpose we will demand of that the coefficients of the zero and first
OBR's must satisfy the condition

\begin{equation} \label{13}
\tilde{\Av}_1(E)=C(E)\cdot\Av_0(E+\Delta),
\end{equation}

\noindent where $C(E)$ is a complex function. Then, by Eq. (\ref{8}),
$\Av_0(E)$ must obey the functional equation

\begin{equation} \label{14}
\Av_0(E)=C(E)Z(E)\Av_0(E+\Delta).
\end{equation}
It is easy to check that $\Av_0(E)$ is determined by this equation with
an accuracy of a scalar periodical function, $\omega (E)$;
$\omega(E+\Delta)=\omega(E)$. Namely, if the function $\Av_0(E)$ is a
solution, so will be the one $\omega(E)\cdot\Av_0(E)$.

Now, taking into account (\ref{13}) and (\ref{14}) in the relation
(\ref{10}) we have

\begin{equation} \label{15}
\tilde{\Av}_N(E)=G_N(E)\Av_0(E+N\Delta),
\end{equation}
\noindent where $G_N(E)=\prod_{n=0}^{N-1}C(E+n\Delta)$.

As in the ZFP \cite{Ch2}, Eqs. (\ref{12}) and (\ref{15}) provide
theoretically the way of deriving, in explicit form, the expressions for
the $N$-barrier transfer matrix (\ref{11}) in terms of $\Av_0(E)$, i.e.
in terms of unit-cell characteristics. However as will be seen from the
following, in this case it gives no preferences in calculating the
${\cal Z}_{(1,N)}(E)$.

Considering (\ref{15}) and the relation
$G_{n+1}(E)=C(E)G_n(E+\Delta)$, one can show that
\[
\tilde{\Av}_{n+1}(E)=C(E)\tilde{\Av}_n(E+\Delta).
\]
Such a connection between the coefficients of two adjacent OBR's
provides fulfilling the symmetry condition (\ref{2}). Namely,

\begin{equation} \label{16}
\Psi_{\cal E}(x+D;E)=\alpha^{-1}(E)C(E)\Psi_{\cal E}(x;E+\Delta).
\end{equation}

So, in the TMM the symmetry condition leads to the functional equation
(\ref{14}) for coefficients of the general solution of the Schr\"{o}dinger
equation.

\section{Solutions of the functional equation}

According to the theory of functional equations \cite{Kuc}, in order to
solve Eq. (\ref{14}) one needs to define the auxiliary functions
$\eta_n(E)$, where $n=0,1,\ldots$, with help of the relationships

\begin{equation} \label{17}
\eta_0(E)=C(E)Z(E)\eta_0(E);
\end{equation}
\begin{equation} \label{18}
\eta_n(E)=C(E)Z(E)\eta_{n-1}(E+\Delta).
\end{equation}
Then the solution of Eq. (\ref{14}) can be written, it is easily
checked, in the form

\begin{equation} \label{19}
\Av_0(E)=\lim_{n\to \infty}\eta_n(E).
\end{equation}
In fact, it means that we have to solve the auxiliary equation (\ref{17})
and to prove the existence of the limit (\ref{19}).

Considering Eqs. (\ref{17}) and (\ref{18}), we can write the limit
(\ref{19}) also as

\begin{equation} \label{20}
\Av_0(E)=G_\infty(E){\cal Z}_{(1,\infty)}(E)\cdot\tilde{\eta}_0,
\end{equation}
where $\tilde{\eta_0}=\lim_{n\to \infty}\eta_0(E+n\Delta).$

The finding of $\Av_0(E)$ is seen to be associated with calculating
the matrix  ${\cal Z}_{(1,\infty)}(E)$ for the semi-infinite structure.
That is the reason that to derive expressions for ${\cal Z}_{(1,N)}(E)$
in terms of $\Av_0(E)$ is of no interest in this approach.

Let us begin with solving Eq. (\ref{17}). It can be rewritten as

\begin{equation} \label{21}
\frac{\eta_0^{(-)}}{\eta_0^{(+)}}=\frac{C^{-1}-q}{p}=
\frac{p^*}{C^{-1}-q^*}; \hspace{8mm}
\eta_0=\left(\begin{array}{c} \eta_0^{(+)} \\ \eta_0^{(-)}
\end{array} \right).
\end{equation}

This equation coincides by form with equation (8) (see the Ref.
\cite{Ch2}) in the ZFP.  The only difference is that the matrix
$Z(E)$ describes the one-cell potential which involves the
electric-field effect. Here the graduation of the energy scale into
Bloch bands ("allowed" energy regions) and gaps ("forbidden" energy
ones) arises as well. But such a division does not yield the energy
spectrum to the given problem and is exclusive of an auxiliary
significance.

Since $det Z(E)=1$ the solutions of the characteristic equation
(\ref{21}) (the right equality) are two reciprocal quantities. In
choosing the required root, for any energy region, we must proceed from
the fact that the function $C(E)$ must have the limit when
$E\to\infty$.  Otherwise, the limit $\tilde{\eta_0}$ does not exist
too, and hence Exp. (\ref{20}) loses its meaning.

Let us show that the solutions of the auxiliary equation (\ref{21}),
having the properties needed, are expressed by

\[
C_1(E)=\frac{1}{q+y};  \hspace{8mm}
\eta_0^{(+)}|_1=1; \hspace{8mm}
\eta_0^{(-)}|_1=\frac{y}{p};
\]

\[
C_2(E)=\frac{1}{q^*-y}; \hspace{8mm}
\eta_0^{(+)}|_2=-\frac{y}{p^*}; \hspace{8mm}
\eta_0^{(-)}|_2=1; \]
\[C_2=C_1^{-1}, \hspace{8mm} y=-\frac{i |p|^2\cdot
sign(u)}{|u|+\sqrt{u^2-|p|^2}}, \hspace{8mm} u=Im(q).\]

\newcommand{\pme}{\stackrel{<}{\sim}}
\newcommand{\ED}{E+N\Delta}
\newcommand{\SS}{S_{E,\Delta}}
\newcommand{\SSf}{S^f_{E,\Delta}}
\newcommand{\SSa}{S^a_{E,\Delta}}
First of all it should be noted that the limit $\tilde{\eta_0}$
is calculated on the set of the equidistant points $E_n$, where
$E_n=E+n\Delta$; $n=0,1,\ldots$. The set will be denoted by $\SS$, in
doing so we emphasize its dependency on the parameters $E$ and $\Delta$.
It is supposed that $E$ varies in the interval $(0,\Delta]$. The set
$\SS$ consists of the two subsets $\SSa$ and $\SSf$ whose points
belong to the bands ($|u|>|p|$) and gaps ($|u|\le|p|$), respectively.
As will be shown below, the behavior of the vector-function
$\eta_0(E_n)$ on the subsets $\SSa$ and $\SSf$ differs qualitatively.
Therefore, there exists no limit, $\tilde{\eta_0}$, when both the
subsets are infinite.

It follows from the general considerations that the number of
points in $\SSa$ and $\SSf$ depends on the bands and gaps width as well
as on their location on the energy scale. As will become clear from the
following, both factors are sufficient to investigate for the
rectangular barrier ($v(x)=v_0$). By using the explicit expressions
for the tunneling parameters of the rectangular barrier (see for
example \cite{Ch1}), one can show that for the matrix element
$\tilde{p}$ in the high-energy region the inequality $|\tilde{p}|\pme
(v_0/2) E^{-1}$ is valid.  The asymptote of the phase $J(E)$ is the
function $k_0(E)d$ ($d=b_1-a_0$ is the barrier width); for the bigger
is the electron energy, the more its motion similar to that of the
free electron.  Thus, in the high-energy region the centres of the
gaps (the points which satisfy the equation $\sin(J(E)+k_0l)=0$ are
meant) for periodical structures formed of the rectangular
barriers asymptotically coincide on the energy scale with the points
$E_L$, where \[E_L=L^2\epsilon; \hspace{4mm}
\epsilon=\frac{\pi^2\hbar^2}{2m D^2}; \hspace{4mm} L=0,1,\ldots.\] The
distance between the gaps centres in the high-energy region is,
consequently, a multiple of the constant $\epsilon$. In this case the
gaps width tends to zero with increasing $E$, and the bands width, on
the contrary, does to infinity (see \cite{Ra2,Ao1,Av2}).  These
findings are not changed by the presence of the step at the right
boundary of the barrier, because the corresponding matrix $\Gamma$ is
real, and, besides, $|p_s(E)|\sim E^{-1}$ as for the rectangular
barrier.  Namely, for the matrix $Z(E)$ we have \begin{equation}
\label{24} |p(E)|\pme \frac{v_0}{2E}, \hspace{8mm} arg(q)\approx
k_0(E)D, \end{equation} here it is also taken into account that
$q=\tilde{q}q_s+\tilde{p}p_s$, $p=\tilde{q}p_s+\tilde{p}q_s$;
$|\tilde{q}|^2-|\tilde{p}|^2=1$, $q_s^2-p_s^2=1$. The asymptotics in
the high-energy region is not changed also when going to the
general-form barrier, because in this case the inequality
$|\tilde{p}(E)|\pme (v_{max}/2)E^{-1}$ holds, and the asymptotics of
$p_s(E)$ remains the same; here $v_{max}$ is the maximum of $v(x)$ by
modulus.

It follows from the above that the subset $\SSa$ is always
infinite, and $\SSf$ is infinite in the exceptional ("resonance") cases
only: $E=\Delta=r\epsilon$, where $r$ is a rational number. There
is no limit $\tilde{\eta_0}$ under these conditions, for the modulus of
the functions $\eta_0^{(-)}(E)|_1$ and $\eta_0^{(+)}(E)|_2$ is equal to
unity on $\SSf$, but on $\SSa$ it varies between the limits zero and
unity.

At the given $\Delta$ the set of the energy values, where the
"resonance" takes place, is no more than a countable set. It is
eventually connected with the fact that the gaps width is zero in the
limit $L\to\infty$. Any arbitrary small variation of $E$ removes the
points $E_n$, beginning with the some number ${\cal N}$, from the gaps.
Since all these points belong to the subset $\SSa$, where the
inequality $|u|>|p|$ holds, there is a such $\delta>0$ that for all
$n>{\cal N}$ the condition $|u(E_n)|\ge|p(E_n)|^{1-\delta}$ is valid
(note, that $|p|^2<1$ in the high-energy region). It follows from here
that at these points we have

\[|y|=\frac{|p|^2}{|u|+\sqrt{u^2-|p|^2}}<
\frac{|p|^2}{|u|}\le|p|^{1+\delta}.\]
And hence
\[|\eta_0^{(-)}(E_n)|_1=\frac{|y|}{|p|}\le|p(E_n)|^\delta\pme
\gamma E_n^{-\delta},\]
where $\gamma=(v_{max}/2)^\delta$. The same asymptotics takes place for
$\eta_0^{(+)}(E_n)|_2$.  It means that almost everywhere on the energy
scale

\begin{equation} \label{25}
\tilde{\eta_0}|_1=\left(\begin{array}{c} 1 \\ 0
\end{array} \right);  \hspace{8mm}
\tilde{\eta_0}|_2=\left(\begin{array}{c} 0 \\ 1
\end{array} \right).
\end{equation}

Now substituting (\ref{25}) into (\ref{20}) we get the final expressions
for two solutions of functional equation (\ref{14}):

\begin{equation} \label{26}
\Av_0^{(1)}=\left(\begin{array}{c} Q_{(1,\infty)}G_\infty \\
P^*_{(1,\infty)}G_\infty \end{array} \right); \hspace{8mm}
\Av_0^{(2)}=\left(\begin{array}{c} P_{(1,\infty)}G^{-1}_\infty \\
Q^*_{(1,\infty)}G_\infty^{-1} \end{array} \right),
\end{equation}
where $G_\infty=G_\infty^{(1)}=1/G_\infty^{(2)}$; $Q_{(1,\infty)}$
and $P_{(1,\infty)}$ are the elements of ${\cal Z}_{(1,\infty)}$.
Then from (\ref{15}) it follows that
\begin{equation} \label{27}
\tilde{\Av}_\infty^{(1)}(E)=\left(\begin{array}{c}
G_\infty(E) \\ 0 \end{array} \right);
\hspace{8mm} \tilde{\Av}_\infty^{(2)}(E)=\left(\begin{array}{c}
0 \\ G_\infty^{-1}(E)
\end{array} \right).
\end{equation}

Expressions (\ref{3}), (\ref{10}) and (\ref{26}) provide two
independent functions $\Psi^{(1)}_{\cal E}(x;E)$ and $\Psi^{(2)}_{\cal
E}(x;E)$. Both solutions are current-carrying. The corresponding
probability flows, $I_{(1)}(E)$ and $I_{(2)}(E)$, are

\begin{equation} \label{28}
I_{(1)}(E)=\hbar m^{-1}k_0(E)|G_\infty(E)|^2; \hspace{4mm}
I_{(2)}(E)=\hbar m^{-1}k_0(E)|G_\infty(E)|^{-2}.
\end{equation}

Now we have to prove that the limit in (\ref{19}) exists. Otherwise
Exps. (\ref{26})-(\ref{28}) are meaningless.

\section{On existence of the solutions \dff}
\renewcommand{\ED}{E+n\Delta}

For a complex-valued matrix $H$ and vector $\Av$ let us define the
norms
\[\| H\|=\max_j\sqrt{\sum_{i=1}^2|h_{ij}|^2}; \hspace{4mm} j=1,2;
\hspace{4mm} \| \Av\|=|A^{(+)}|+|A^{(-)}|.\]
In particular, it means that $\|Z\|^2=1+2|p|^2$.  Considering the
first solution we will prove that for any given $E$ and $\varepsilon>0$
we can find such number $N$ that

\begin{equation} \label{29}
\|\eta_n(E)-\eta_{n-1}(E)\|<\varepsilon
\end{equation}
 for $n>N$. Since \[\eta_n(E)=G_n(E){\cal
Z}_{(1,n)}(E)\eta_0(E_n),\] (here $E_n=E+n\Delta$) we have
\begin{equation} \label{30}
\|\eta_n(E)-\eta_{n-1}(E)\|\le |G_{n-1}(E)| \cdot\|{\cal
Z}_{(1,n-1)}(E)\|\cdot F(E),
\end{equation}
where $F(E)=\|C(E_n)Z(E_n)\eta_0(E_n)-\eta_0(E_{n-1})\|$.

Let us show that the first two norms are bounded as $n\to\infty$.
We have \[|G_\infty(E)|^{-1}=\prod_{n=0}^{\infty}|C(E_n)|^{-1}=
\prod_{n=0}^\infty|q(E_n)+y(E_n)|\le\]
\begin{equation} \label{31}
\le\prod_{n=0}^{\infty}|q(E_n)|\cdot\Biggl(1+\frac{|y(E_n)|}
{|q(E_n)|}\Biggr).
\end{equation}

The convergence of both norms in (\ref{31}) is equivalent to that of
the series $\sum_{n=0}^{\infty} n^{-2}$, because $|q|=\sqrt{1+|p|^2}$,
$y\sim |p|^2$, $|p|\sim n^{-1}$. Since this series converges, the
infinite product $|G_\infty(E)|$ does as well.

For the matrix describing the semi-infinite structures we have
\[\|{\cal Z}_{(1,\infty)}(E)\|^2\le\prod_{n=0}^{\infty}\|Z(E_n)\|
=\prod_{n=0}^{\infty}(1+2|p(E_n)|^2).\] Obviously, this product
converges for the same reason as in (\ref{31}). In addition, since
\[\|{\cal Z}_{(1,\infty)}(E)\|^2\equiv
1+2\frac{R_{(1,\infty)}(E)}{T_{(1,\infty)}(E)},\]
we have that $T_{(1,\infty)}(E)\ne 0$. So that the semi-infinite
structure must not be absolutely opaque for an electron.

Now, it remains to be shown that $F$ in (\ref{30}) approaches zero with
increasing $n$. Using (\ref{17}) we have

\begin{equation} \label{32}
F=\|\eta_0(E_n)-\eta_0(E_{n-1})\|\pme 2\gamma n^{-\delta}.
\end{equation}

Since the norms $|G_\infty(E)|$ and $\|{\cal Z}_{(1,\infty)}(E)\|$
are bounded there is  \[\max_j (|G_{j}(E)| \cdot \|{\cal
Z}_{(1,j)}(E)\|),\] where $j=1,2,\ldots$. Together with (\ref{32}) this
provides fulfilling the inequality (\ref{29}), that proves the existence
of the limit in (\ref{19}). For the second solution the arguments are
similar.

\section{Conclusions}

At first glance the functions \dff may be calculated by this
method in the region located to the right of the zero cell
only.  However it should be noted that any unit cell of the periodical
structure may be taken as a zero cell. Then by making use of the
transfer matrix which connects solutions in the zero cell and in the
regions to the left of it, one can calculate the functions \dff on
the whole axis $Ox$.

Since both functions \dff are current-carrying they increase infinitely
by modulus in the classically inaccessible range when $x\to -\infty$,
according to the general properties of the one-dimensional
Schr\"{o}dinger equation.  Thus, for infinite structures, \dff are no
solutions to the problem.  However, for any $E$ (excluding a countable
set for certain values of $\Delta$), the solution (undegenerate) for
the infinite structure can be obtained as a linear combination of these
functions.  As a result we arrive at two important conclusions.  First,
for the limited periodical potentials the energy spectrum of an
electron in the problem for infinite structures is continuous (so that
the Wannier-Stark states, by the model, may exist only as the
quasi-stationary ones).  Second, the stationary wave functions of an
electron in the infinite structures, being the linear combinations
of the functions \dff, do not satisfy the symmetry condition (\ref{2}).
(There exist a mistaken opinion that the continuity of the spectrum in
this problem is obvious.  The following arguments are used in this
case.  Namely, the energy spectrum is continuous since

a) the range, where $x$ is large, is classically accessible for an
electron;

b) the periodical potential is negligible in comparison with the Stark
potential when $x\to \infty$, and, consequently, the electron motion in
this range is of the free electron type (see, for example, \cite{Wan}).
However, it should be noted that the first statement is valid only if
$V(x)$ remains finite at the plus infinity. But if $V(x)\to -\infty$ as
$x\to \infty$, then the electron spectrum may be both continuous and
discrete, depending on the monotonicity and rate of decreasing
$V(x)$ at $x\to \infty$ (see, for example, \cite{Tic}). It follows also
from this the erroneousness of the second argument, because on the
whole axis $OX$ the derivatives of $V(x)$ (and, hence, its
monotonicity) are determined by the periodical component of the
potential.)

It is interesting also to dwell on the question of the connection of the
given problem to the ZFP. We will start with the fact that the wave
functions \dff are defined in terms of the solutions of auxiliary
equation (\ref{21}) describing formally the electron motion in the
periodic structures in the absence of an electric field. In addition,
for finite structures the functions \dff, by their properties, are
close to the solutions of the ZFP, if $N\Delta\ll E$ ($N$ is the number
of unit cells in the structure). In particular, if values of $E$ are in
the band, then \dff, in the given interval, are close to the
usual Bloch functions. It is the case when an electrical field has a
weak effect on the electron with the energy $E$. However there is no
transition from the given problem to the ZFP when the periodical
structure is considered on the whole axis $Ox$. The wave functions \dff
are unbounded, when $x\to -\infty$, at any value of the electric-field
strength.

Some comments should be further made about the role of the
Zener tunneling (ZT) which have been the subject of great interest (see,
for example, \cite{Ro2} and referrers therein) since paper
\cite{Zen}. Strictly speaking, by this concept is meant the electron
transitions between the bands, and therefore the latter relates to the
non-stationary case. In the models with the vector potential the Zener
tunneling is caused by the accelerating effect of the field,
resulting in that a Bloch electron passes (tunnels) from the
lower bands to the upper.
In our approach we investigate stationary states. Nevertheless, we can
draw some conclusions on the question. It is possible because symmetry
condition (\ref{16}), governing the functions \dff, links their $E$-
and $x$-dependencies. In particular, for \dff relationship
(\ref{15}) is valid. Note also that $\tilde{\Av}_\infty$ (see
(\ref{27})) is a bounded non-zero value. It provides the asymptotics
$\Av_n\sim n^{-1/4}$ and $\Av_0(E)\sim E^{-1/4}$ (as for Airy's
functions).  Thus the probability that an electron is in the $n$-th
unit cell or it has the energy $E$ decreases with increasing of these
parameters by the power law instead of the exponential one. This result
makes the conclusion presented in Ref. \cite{Ao1} more precise.

Also it follows from the above that the well-known Bloch
oscillations can exist only as the decaying ones. As for the
experimental evidences of the long-lived Bloch oscillations and the
Wannier-Stark ladders in superlattices, it is not the question of
correctness of our approach. This implies only that one needs to find
the mathematical model which would be more suitable for the experiments
on superlattices. In the following paper we are going to present such a
model.

\section*{Acknowledgment}
The author thanks professor G.F.Karavaev for useful discussions.

\end{document}